\documentclass[onecolumn,iop,apj]{emulateapj}   

 \usepackage[T1]{fontenc}
\usepackage{graphicx}        
\usepackage{amssymb}         
\usepackage{amsmath}         
\usepackage[usenames,dvipsnames]{color}          
\usepackage{epstopdf}
\usepackage{float}
\usepackage[normalem]{ulem}
\definecolor{darkblue}{rgb}{0,0,0.5}
\usepackage{bm}
\DeclareFontFamily{OT1}{pzc}{}
\DeclareFontShape{OT1}{pzc}{m}{it}%
             {<-> s * [1.1500] pzcmi7t}{}
\DeclareMathAlphabet{\mathscr}{OT1}{pzc}%
                                 {m}{it}

\DeclareGraphicsExtensions{.pdf,.png,.jpg,.tif,.pdf}

\usepackage{amsmath}

\begin{document}

\title{MHD wave refraction and the acoustic halo effect around solar active regions - a 3D study}
\author{Carlos Rijs}
\author{Hamed Moradi}
\author{Damien Przybylski}
\author{Paul S Cally}

\affil{Monash Centre for Astrophysics and School of Mathematical Sciences,\\ Monash University, Clayton, Victoria 3168, Australia}
  \email{carlos.rijs@monash.edu}


\begin{abstract}
An enhancement in high-frequency acoustic power is commonly observed in the solar photosphere and chromosphere surrounding magnetic active regions. We perform 3D linear forward wave modelling with a simple wavelet pulse acoustic source to ascertain whether the formation of the acoustic halo is caused by MHD mode conversion through regions of moderate and inclined magnetic fields. This conversion type is most efficient when high frequency waves from below intersect magnetic field lines at a large angle. We find a strong relationship between halo formation and the equipartition surface at which the Alfv\'en speed $a$ matches the sound speed $c$, lending support to the theory that photospheric and chromospheric halo enhancement is due to the creation and subsequent reflection of magnetically dominated fast waves from essentially acoustic waves as they cross $a=c$. In simulations where we have capped $a$ such that waves are not permitted to refract after reaching the $a=c$ height, halos are non-existent, which suggests that the power enhancement is wholly dependent on returning fast waves. We also reproduce some of the observed halo properties, such as a dual 6 and 8 mHz enhancement structure and a spatial spreading of the halo with height.

\end{abstract}

\keywords{magnetohydrodynamics (MHD) -- Sun: helioseismology -- Sun: magnetic fields -- Sun: oscillations -- waves}

\section{INTRODUCTION}

At present the local behaviour of wave modes in and around solar regions of significant magnetic field strength is not fully understood (see \citealt{gizon2010} and \citealt{moradi2010} for recent reviews). Strong field regions alter the characteristics of otherwise simple acoustic waves, resulting in vastly different behaviour for waves on either side of the fast/slow magneto-acoustic branch. The process of mode conversion has a significant effect on these incoming waves. As initially suggested by \citet{spruit1992}, conversion from fast waves to downward travelling, field aligned slow waves has been shown to be the likely responsible mechanism \citep{cally1997,cally2003} for the well observed umbral $p$-mode absorption and associated phase shifts \citep{braun1987}. \\
The physical mechanism behind the power enhancement observed around sunspots and active regions at frequencies above the local acoustic cutoff (the so called acoustic halo) is still unknown.\\
First observation of the halo began several years after the discovery of the aforementioned $p$-mode absorption. \citet{braun1987}, assessing the viability of acoustic power maps as a diagnostic tool for local helioseismology, noticed not only the well observed reduction of power at around 3 mHz, but also a high frequency enhancement at around 6 mHz extending many arcseconds radially. The discovery was soon verified by \citet{brown1992} and by \citet{toner1993}. With the increasing observational accuracy made available with the \emph{Solar and Heliospheric Observatory} (SOHO), it became clear that an enhancement was present only in the power of velocity amplitudes and not in measurements of the continuum intensity \citep{hindman1998,jain2002}.  \\
More recently, \citet{schunker2011} showed that halo power excess is prominent for moderate strength ($150$ G $< |\mathbf{B}| < 800$G) and near horizontal field, and that, importantly, the peak halo frequency $\nu$  increases in proportion to the field strength.\\
\citet{rajaguru2012}, using the \emph{Helioseismic and Magnetic Imager} (HMI) and \emph{Atmospheric Imaging Assembly} (AIA) instruments onboard the \emph{Solar Dynamics Observatory} (SDO), have calculated doppler velocities and intensities corresponding to observables at heights of between 0-430 km above the photosphere (where the continuum optial depth is unity). The most important findings can be summarised as follows: \\
\\
1. The standard $5.5-7$ mHz doppler velocity halo observed by all above references is clearly visible and is strongest in near-horizontal field regions, decreasing in amplitude as the field becomes more vertically aligned. As the field strength increases, the halo peaks at a greater frequency ($\nu$). \citep{schunker2011}. \\
\\
2. The presence of the halo is extremely dependent on height. At the base of the photosphere ($z=0$) for weak field regions, there is a uniform wave power above the acoustic cutoff, which is to be expected. However, at heights of around $140$ km (corresponding to HMI observations of the Fe 6173.34 {\AA} line) the situation is markedly different, and the enhancement (with respect to quiet Sun values) comes into effect. \\
\\
3. The enhancement is also visible in the chromosphere in intensities. A Fourier analysis of the AIA intensity data for the 1600 {\AA} and 1700 {\AA} wavelength channels (corresponding approximately to average heights of 430 and 360 km respectively) shows high frequency enhancements, which spread with height \citep{rajaguru2012}. Once again, no enhancement is present in the continuum intensity power (taken from a height of around $z=$ 0 km)\\
\\
4. There is a secondary halo which exists in the 7.5-9.5 mHz range which is only manifested amongst stronger horizontal fields ($|\mathbf{B}|>300G$). Such conditions likely correspond to a canopy field surrounding a strong sunspot, and this higher frequency halo is shown in power maps to be highly localized spatially. Radially outwards from this higher $\nu$ field is a region of significantly reduced power, which in turn is surrounded by a diffuse, weak halo region, extending radially many Mm into relatively quiet regions \citep{rajaguru2012}.\\
\\
There have been several recent theories attempting to explain the acoustic halo. Early theories suggested that the halos correspond to areas of increased acoustic emission \citep{brown1992}. However enhancements are generally not observed in continuum intensity \citep{rajaguru2012}, which casts doubt on this hypothesis.\\
\citet{jacoutot2008} has performed radiative MHD simulations with an emphasis on the effect of the magnetic field on the frequencies of excitation originating from the solar convection zone. They found that the field shrinks the granulation scale size and shifts the local oscillation frequency upwards to higher values, consistent with halos. \\
\citet{kuridze2008} shows analytically that it is possible for waves of azimuthal wave number $m>1$ to become trapped under small canopy field regions, while \citet{hanasoge2009} suggests that the local oscillation is shifted preferentially from high to low mode mass \citep{bogdan1996} due to the flux tube acting as a wave scatterer, and that essentially mode energy is being reorganised more significantly for high frequency waves because of their greater propensity for scattering. \\
Additionally the overlying magnetic canopy itself has been shown to be linked with photospheric power enhancement by \cite{muglach2005}, and in particular the mode conversion process that is governed by the ratio of the Alfv\'en speed ($a$) and the sound speed ($c$).  \\
We intend to follow up on the initial simulations and theory of \citet{khomenko2009}, who have suggested that fast/slow mode conversion and transmission may be the dominant mechanism behind halo formation. \\
\citet{khomenko2009} have performed 2D wave propagation simulations through a magneto-hydrostatic sunspot atmosphere with a wavelet source and observed a power enhancement of around 40-60\% in acoustic power with respect to the quiet sun. The enhancement also correlates well with the $a=c$ equipartition region for the photosphere (where they have defined the photosphere as the height at which the optical depth scale is unity).\\
The results suggest that the halos could occur simply as the addition of energy from high frequency non-trapped waves which have travelled above $a=c$ and undergone mode conversion. In this instance, when $a$ and $c$ coincide, the normally separate branches of magnetoacoustic waves (the fast and the slow wave) may interact. If the upcoming waves are of sufficiently high frequency to penetrate above the acoustic cut-off and travel into the $a>c$ atmosphere, the wave's energy will be partially re-assigned into the fast or slow mode depending on the relationship between the wavevector and the orientation of the magnetic field (See \citealt{cally2006,schunker2006} for details or \citealt{cally2007} for an easily accesible review). The slow waves are strongly field aligned at small $\beta$ and may contribute to the diffuse, spatially extended halos observed by \citet{rajaguru2012}.\\ 
The fast waves however will refract due to the rapidly increasing profile of $a$, and eventually reflect where $\omega^{2}=a^{2}k_{h}^{2}$ (where $\omega$ is the angular frequnecy and $k_{h}$ is the horizontal component of the wavenumber), depositing energy into regions below. Low frequency power should therefore not be enhanced in any way, as these waves are unable to reach such heights, except in special wave-field configurations where the ramp effect may reduce the effective acoustic cutoff frequency \citep{cally2006,schunker2006,cally2007}. \\
The fast-acoustic-to-fast-magnetic conversion is favoured at a higher frequency and more importantly by a large attack angle between the incoming wave and the magnetic field lines, which potentially explains why halos are consistently observed at near horizontal fields (i.e. the line of sight component of velocity makes a large attack angle with a horizontal field).
The mechanism also naturally explains the spreading of the halo that is observed with height \citep{braun1992,brown1992}, given that the $a=c$ height is located further outwards radially from sunspot center as a function of height.\\
Recently \citet{kontogiannis2014} performed an interesting observational study by examining photospheric and chromospheric power enhancements as functions of mode conversion parameters, such as the attack angle. They discovered chromospheric slow wave signatures corresponding to waves following the field lines upwards, as well as reflected fast wave signatures correlating with power enhancement regions for high frequency waves, lending further weight to the importance of mode conversion in this instance.

\section{THE SIMULATION}\label{sec:maths}
\begin{figure*}
\centering
\includegraphics[width=1.0\textwidth,trim=4.5cm 0cm 4.5cm 2cm]{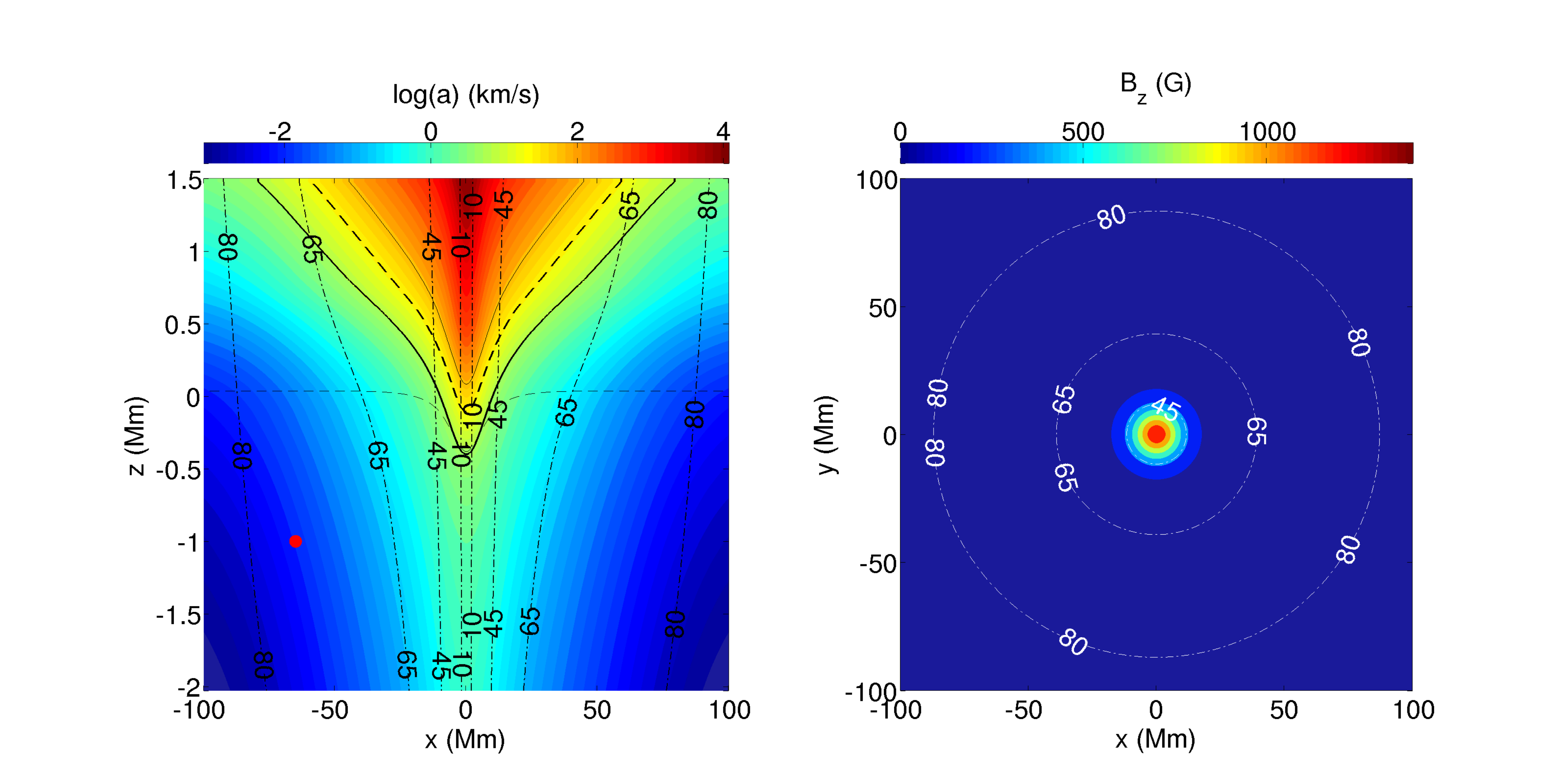}
\caption{The left panel shows A 2D contour cut through the MHS sunspot atmosphere. Vertical dash-dotted curves are contours of field inclination (from the vertical). The thin black dashed curve corresponds to our reference photosphere, where $\log(\tau_{5000})=0$, with a Wilson depression of around 450 km. The thick solid curve is the $a=c$ equipartition height and the thin solid curve is the height at which $a = 80$ km/s. This is the height at which we have capped $a$. The thick dashed curve corresponds to the fast wave turning height for a 6.5 mHz magnetoacoustic wave with $k_{h} \approx 1.5$ Mm$^{-1}$. The background colour contour is $\log(a)$ in km/s. The dot corresponds to the location where the wave source was initiated. Note the highly stretched aspect ratio of the figure, with the abscissa spanning 200 Mm and the ordinate spanning only 3.5 Mm.\\
The right panel shows a contour plot of the vertical component of the magnetic field - $B_{z}$ (in G) for the spot taken at the surface, along with various field inclination contours (dashed contours).}\label{sunspot}
\end{figure*}

\begin{figure*}
   \centering
    \includegraphics[width=1.0\textwidth,trim=4cm 1cm 4cm 2cm]{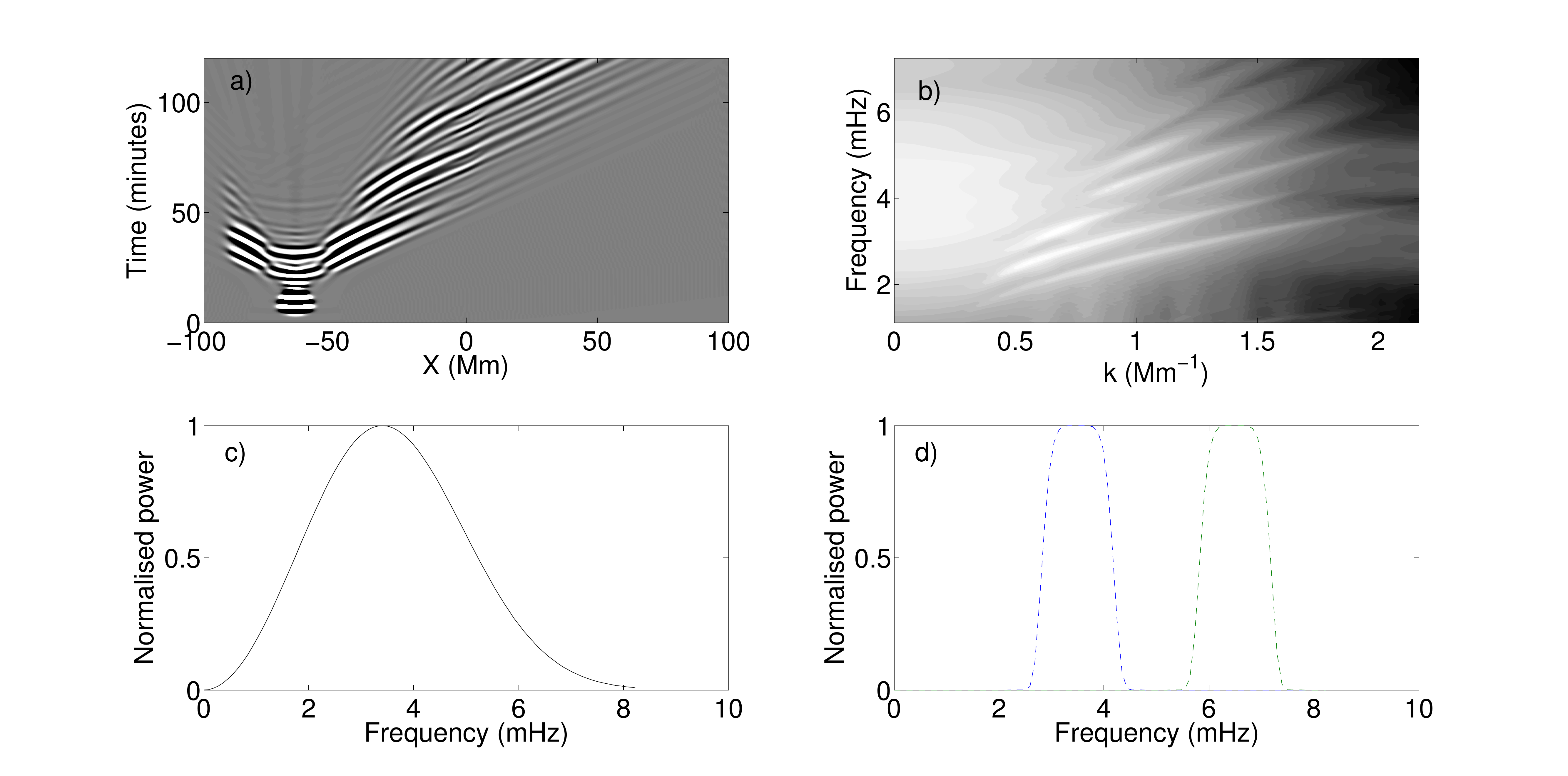}
    \caption{Panel a) shows the time-distance diagram of $v_{z}$, taken along the $\log(\tau_{5000})=0$ contour corresponding to the surface, through the centre of the sunspot. b) shows the full 2-hour azimuthally averaged surface power spectrum of the simulation.  c): The frequency distribution of the wave source, centred at 3.3 mHz. d): The filtering functions used to isolate power at 3.5 and 6.5 mHz.}\label{wave}
\end{figure*}

We proceed by performing 3D simulations in the spirit of \cite{khomenko2009}. The goal is to underake forward modeling of a simple wavepacket propagating through a sunspot-like magnetic field and atmosphere in full 3D and observe the structure of any resultant enhancements in time averaged acoustic power at high frequencies, both spatially and spectrally. In particular we wish to determine whether there is a correlation between the halo formation and the region most important to mode conversion - the $a=c$ equipartition height.\\
Halo formation is manifested on the Sun as an enhancement of time-averaged acoustic power (at frequencies above the acoustic cutoff) at near horizontal magnetic field inclinations with respect to the normal quiet-Sun values. We therefore perform quiet and magnetic simulations separately. After the simulations are complete, the time averaged power at each point of the magnetic simulation can be compared to the corresponding quiet point and regions of enhancement can be identified. In this paper, we shall focus on the power of the vertical component of the velocity perturbation ($v_{z}$), which corresponds observationally to the line-of-sight component of velocity when observing at disk centre.\\
For our quiet-Sun simulations we have used a convectively stabilised version of the Model S \citep{dalsgaard1996} joined onto a VAL-C chromosphere \citep{vernazza1976} modified to minimize convective instabilities which do not lend themselves well to linear wave simulations \citep{parchevsky2007}.
For our magnetic wave propagation simulations, we use the magneto-hydrostatic (MHS) sunspot model of \citet{khomenko2008}, who have joined the self-similar lower photospheric sunspot model of \citet{schluter1958} and \citet{low1980}, with the pressure distributed model of \citet{pizzo1986} to create a consistent magnetic atmosphere. The model has been further enhanced to provide a consistent MHS structure and a convectively stable stable pressure, density and temperature profile, accurate to empirical models of the solar photosphere (Przybylski. 2014 - in preparation).\\
We use the 3D wave propagation code SPARC \citep{hanasogephd2007, hanasoge2007}. The code solves the ideal linearised Eulerian MHD equations in cartesian coordinates for a given magnetic or quiet (non-magnetic) atmosphere. The code is ideally suited to the forward modelling of adiabatic wave propagation, with output consisting of the linear perturbations to the background states of the pressure ($p$), density ($\rho$), magnetic field ($\mathbf{B}$) and vector velocity ($\mathbf{v}$).\\ 
The computational box consists of square, horizontal dimensions $L_{x}=L_{y}=200$ Mm, corresponding to 256 grid points and yielding a horizontal resolution of $dx=0.78125$ Mm. We define a reference photosphere as the height at which $\log(\tau)=0$ (where $\tau$ is the optical depth scale, as calculated from the known thermodynamic values at every point in the box), and the vertical dimension spans from 10 Mm below this surface to 1.85 Mm above it. The vertical axis is scaled in inverse proportion to the sound speed with 215 grid points, yielding grid spacing on the order of $20$ km above the surface, to spacings of around $100$ km at the bottom of the box.\\
In an attempt to keep the simulation as simple as possible and avoid any periodicity, we have implemented sponges along the sides of the box and perfectly matched layers (PML) along the top and bottom. The intention is to ensure that all outgoing waves are damped completely. The top PML takes effect over the top 15 grid points, resulting in a useable box top of 1.5 Mm\\
The most immediate problem when initiating forward modelling in a stratified magnetic atmosphere is the rapid increase with height of  $a$ (where $a=B /\sqrt{4 \pi \rho(z)}$, with $B=|\mathbf{B}|$) caused by the swiftly decreasing profile of $\rho$. The CFL constraint on any explicit finite differences scheme requires that the maximum stable time step at any given point is inversely proportional to the local velocity scale there. In other words, if we extend the box too far into the atmosphere, the timestep required becomes impractically small. Methods of artificially capping $a$ at some manageable value have been well described \citep{hanasoge2012,rempel2009}. We use the method of \citet{rempel2009}. When $a$ begins to dominate over $c$, or in other words the plasma $\beta$ becomes sufficiently small according to a chosen criteria, the limiter will take effect and prevent any further rise in $a$. It is important therefore to ensure that such a limiter takes effect well above the fast wave reflection heights for any high frequency waves of interest (see \citealt{moradi2014} for a discussion on how Alfv\'en limiters may affect the helioseismology). \\
For our simulation we cap $a$ at 80 km/s, yielding a fairly manageable simulation time step of $0.2$ s for our sunspot, which has a peak surface field of $1.4$ kG. \\
A contour plot giving an overview of the magnetic atmosphere that we have chosen to use is shown in Figure \ref{sunspot}.\\
Note that the fast wave reflection height for a typical halo frequency wave is crucially between the $a=c$ layer and the $a=80$ km/s contour, meaning that the primary body of fast waves which undergo mode conversion are free to reflect back downwards unhindered by our upper atmospheric meddling.\\
To begin, we set off a time dependent pulse (essentially adding a source term on the right hand side of the momentum equation) similar to those used by \citet{moradi2014} and \citet{shelyag2009}, of the form
\begin{equation}
v_{z}=\sin\frac{2 \pi t}{p} \exp(-\frac{(t-t_{0})^{2}}{\sigma_{t}^{2}})\exp(-\frac{(\mathbf{x}-\mathbf{x_{0}})^{2}}{\mathbf{\sigma_{x}}^{2}}),
\end{equation}
where $v_{z}$ is the perturbation to the vertical component of the velocity, $p=300s$, $t_{0}=300s$, $\sigma_{t}=100s$ and $\mathbf{x}=(x,y,z)$. $\mathbf{\sigma_{x}}$ was chosen to give a pulse size of approximately 5 grid points in the $x$, $y$ and $z$ directions, with $\mathbf{x_{0}}=$($-65$, $0$, $-1$) Mm. In other words the wave source is located 65 Mm from the sunspot umbra (which lies at ($x$,$y$)$=$($0$,$0$)), in the $y$-plane cutting through the centre of the spot, and at a depth of $1$ Mm below the surface (this position is shown by the red dot in Figure \ref{sunspot}).\\
Such a pulse excites waves with a range of temporal frequencies around 3.3 mHz in somewhat of an approximation to the spectrum observed on the solar photosphere.\\
By utilising such a simple wave source, we may follow a wavepacket as it propagates from the quiet-Sun, through to the magnetically dominated regions of the the sunspot, and analyze any dynamical features (such as halos) as they appear. \\
Our primary goal was to analyse the power distribution on the near side of the sunspot (to the source). As such, the simulation was run for 2 solar hours, which is a sufficient time for the wavepacket to pass through the sunspot umbra.  Figure \ref{wave} shows a summary of the simulation, including a power spectrum and time distance diagram for the 2 hour duration, as well as the frequency distribution of our source function and the associated frequency filters which were applied to the resultant power. \\
From panel a), it is clear that the sponges have been reasonably sucessful at damping the waves, however there is a small amount of reflection occuring off the left hand side sponges, resulting in some very small amplitude waves returning through the simulation domain. The fuzzy region observed at low wavenumber in b) results from wave interactions with the top and bottom box PML. d) gives an example of the filters we apply to the output to isolate specific frequency ranges. \\

\section{RESULTS}\label{sec:results}
\begin{figure*}
   \centering
    \includegraphics[width=1.0\textwidth,trim=0cm 1cm 0cm 1cm]{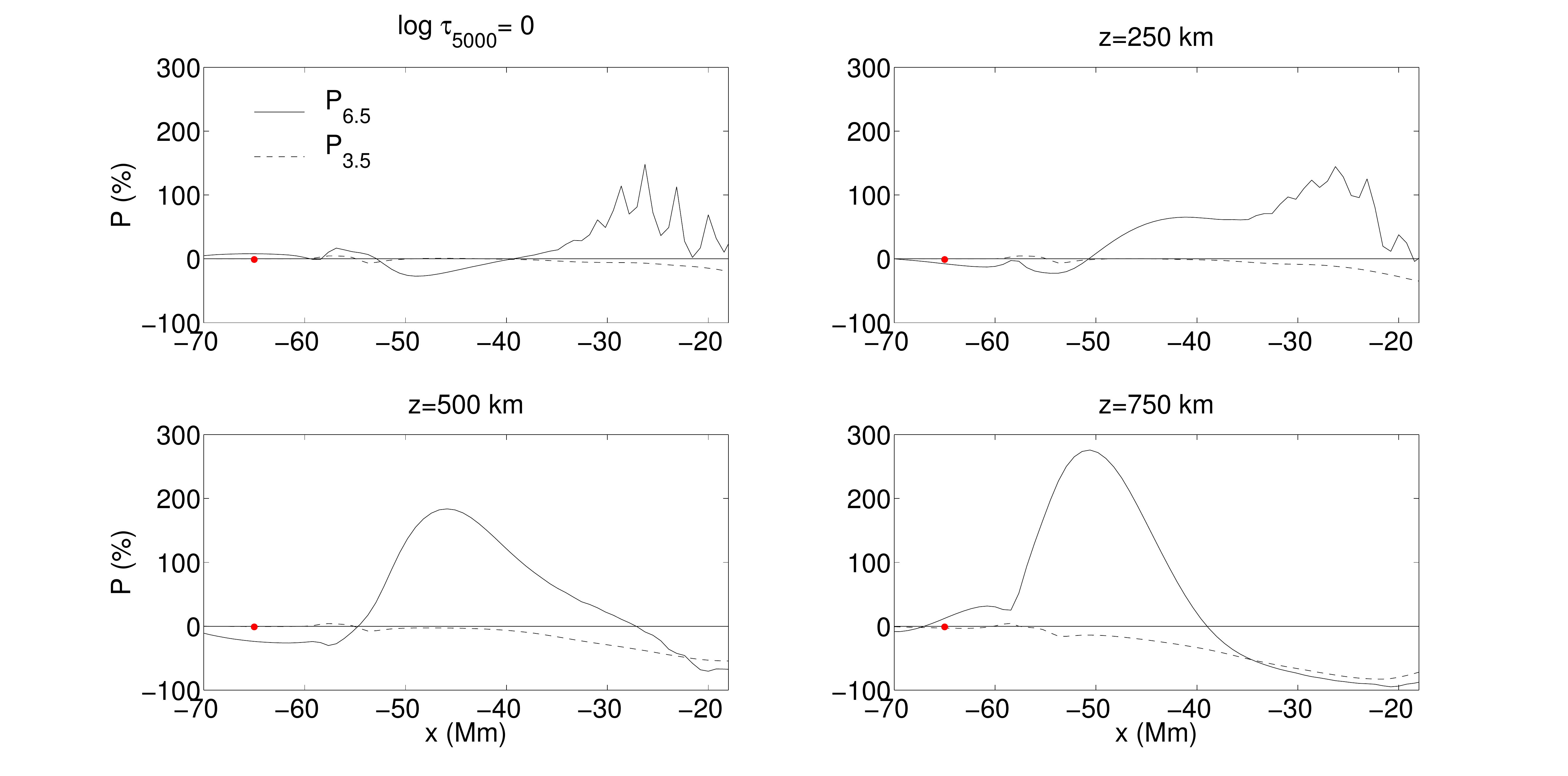}
    \caption{$P_{\nu}$ at the surface, and at constant geometrical heights of $z= 250$, $500$ and $750$ km, extending from around 20 to 70 Mm from the centre of the sunspot umbra. The solid line is the 6.5 mHz enhancement and the dashed line is the 3.5 mHz enhancement. The horizontal solid line indicates $P= 0\%$, which is a quiet sun value. Anything above this line we consider to be an enhancement. The source radial position of -$65$ Mm is shown by the red dot (although it is of course located 1 Mm below $z= 0$).}\label{sp1}
\end{figure*}

\begin{figure*}
   \centering
    \includegraphics[width=1.0\textwidth,trim=8cm 0cm 8cm 1cm]{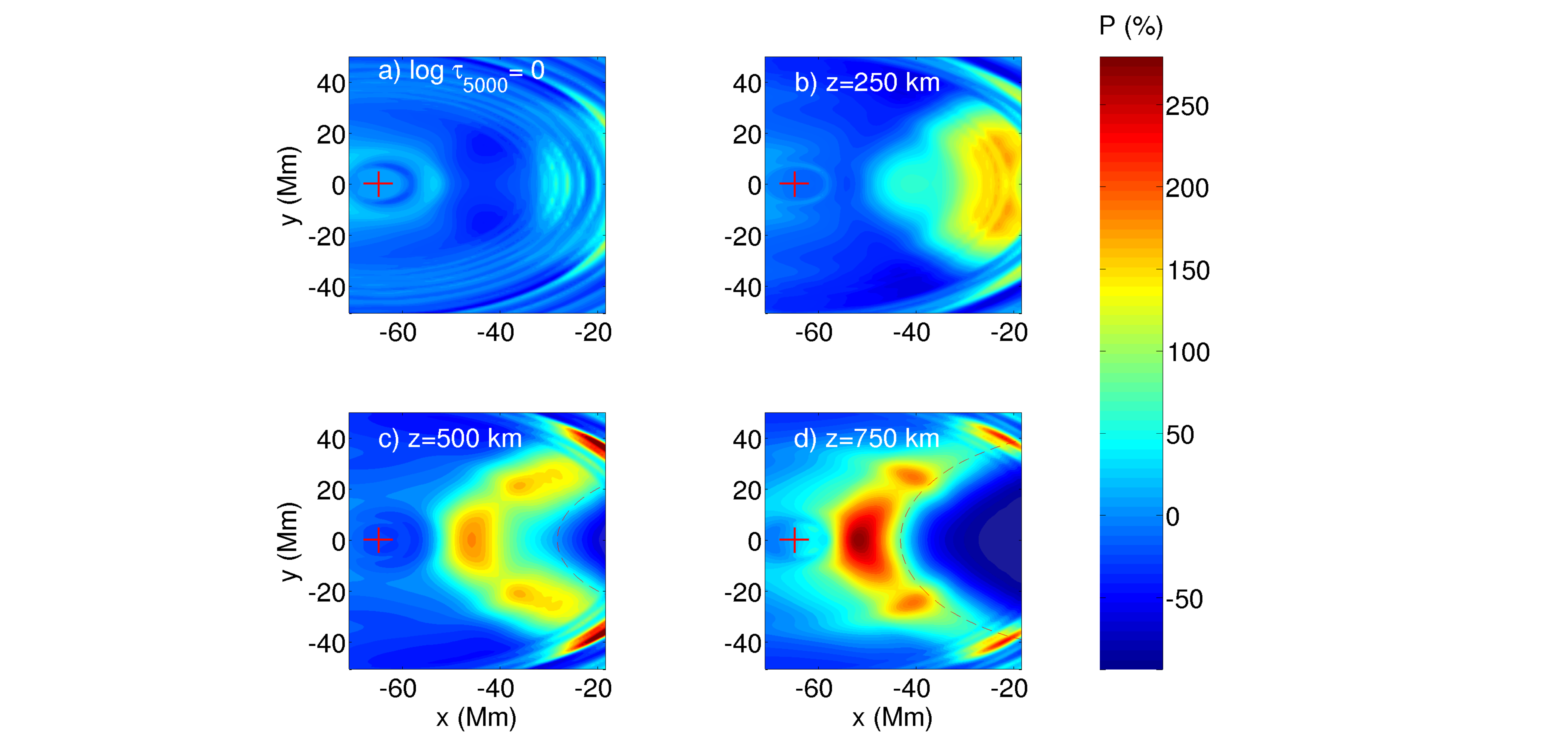}
    \caption{Power maps of $P_{6.5}$ at the same four heights for the $x$-$y$ plane, restricted to the near side of the sunspot. The dashed curve visible at $z=$ 500 km and $750$ km is the $a=c$ contour for that particular height. The source radial position of -65 Mm is shown by the cross }\label{pxy1}
\end{figure*}

The main quantity which we shall use throughout the paper to denote a halo enhancement is $P=(P_{mag}-P_{quiet})/P_{quiet}$. $P_{mag}$ is the 2-hour averaged Fourier power of $v_{z}$ calculated at each point in the sunspot simulation at various heights. $P_{quiet}$ is the analogue to $P_{mag}$ for a completely separate quiet sun simulation. $P$ is therefore the fractional enhancement in power for the sunspot simulation (with respect to the power from the quiet sun simulation) at any grid point. This power can then of course be filtered to isolate particular frequency ranges of interest (denoted as $P_{\nu}$), prior to averaging over the remaining frequency domain.\\
We firstly identify that halos do indeed occur in our simulation. Figure \ref{sp1} shows cuts through spot centre at 4 different heights, with $P_{\nu}$ plotted as a function of radial distance from the centre of the sunspot. We also show power maps for $P_{6.5}$ over the full $x-y$ plane in Figure \ref{pxy1} for the same choice of heights.\\

\begin{figure*}
\centering
    \includegraphics[width=1.0\textwidth,trim=0cm 1cm 0cm 1cm]{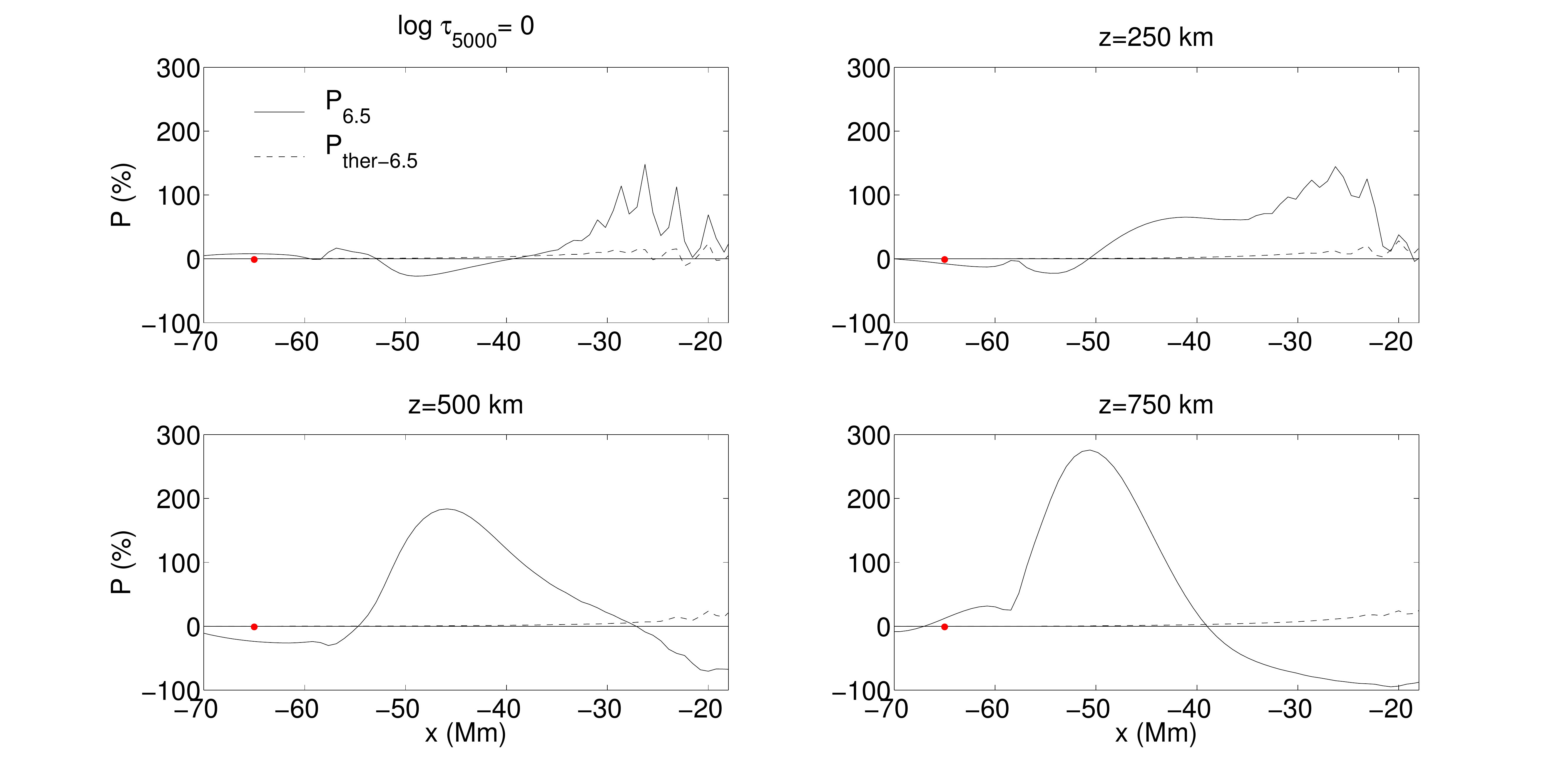}
    \caption{Similarly to Figure \ref{sp1}, we take cuts at 4 heights, including the surface. The solid curve is $P_{6.5}$, just as in Figure \ref{sp1}. The dashed curve is $P_{\text{ther}}$ at 6.5 mHz.}\label{sp2}
\end{figure*}

\begin{figure*}
\centering
    \includegraphics[width=0.99\textwidth,trim=6cm 0cm 6cm 1cm]{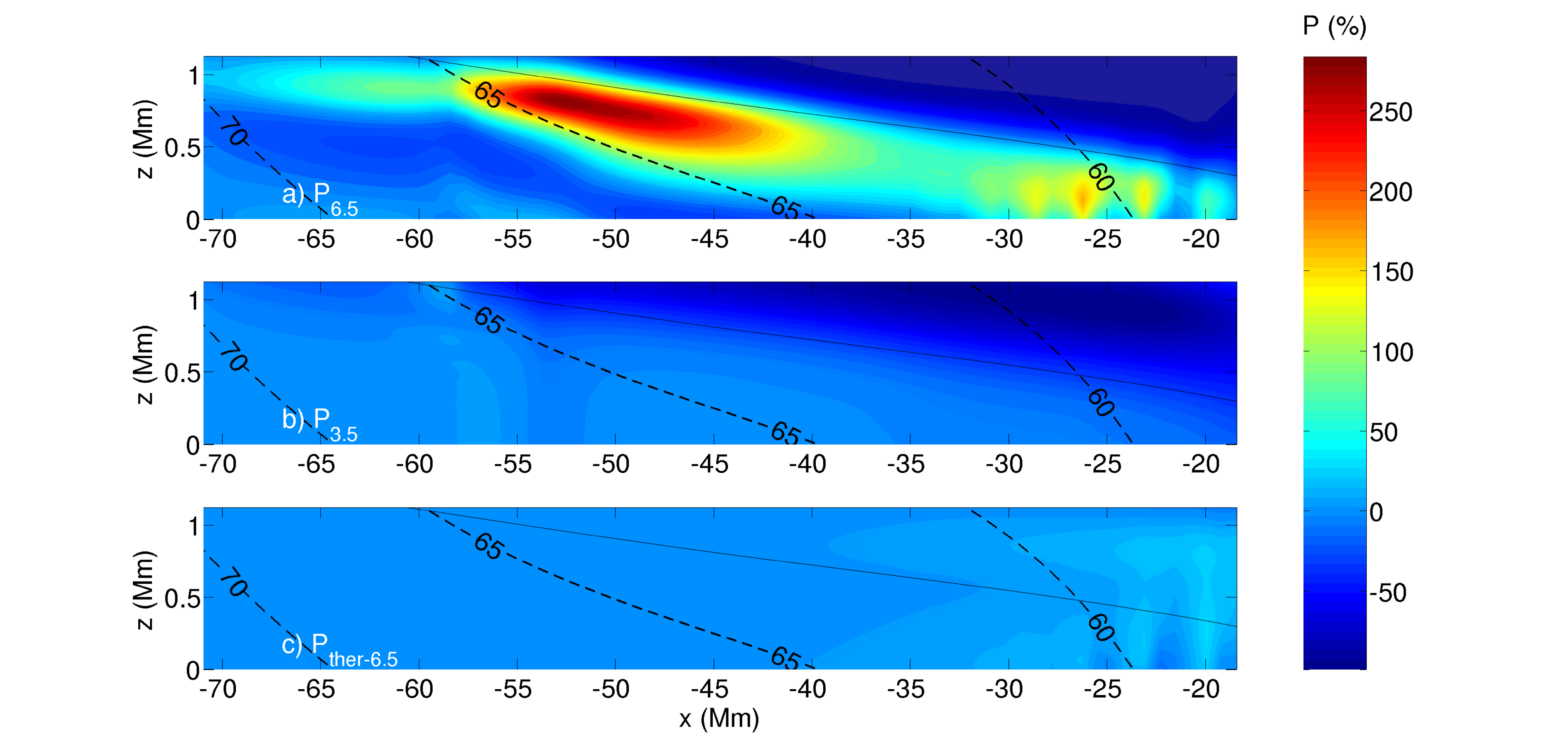}
    \caption{ $P$ and its relationship with the $a=c$ height. This is the vertical plane taken along the line $y=0$ in figure \ref{pxy1}. Panels a) and b) are $P_{6.5}$ and $P_{3.5}$. Panel c) shows the enhancement for the thermal case, $P_{\text{ther}-6.5}$ . The black solid curve is the $a=c$ height and the thick dashed labelled curves are contours corresponding to magnetic field inclinations of 70, 65 and 60 degrees (from left to right) with respect to the vertical.}\label{pxz1}
\end{figure*}

The top left panels correspond to the optical height unity, as shown in Figure \ref{sunspot}, whereas the others are at constant geometrical heights, as labelled.\\
We have chosen to filter the power around both 3.5 and 6.5 mHz, using the filters shown in Figure \ref{wave} d). These frequencies are indicative of low frequency trapped waves (which generally cannot penetrate above the acoustic cut-off height to contribute to halos) and high frequency waves (which will undergo mode conversion and contribute to an enhancement) respectively. We of course expect to see the well observed $p$-mode absorption for the low frequency 3.5 mHz waves, which are suppressed in power when propagating through regions of high magnetic field strength. This can clearly be seen in Figure \ref{sp1}; $P_{3.5}$ shows a deficit with increasing proximity to the sunspot and no power enhancement whatsoever. \\
In contrast, there is a strong 6.5 mHz power enhancement visible at all heights. To be clear, a value of $P=100\%$ indicates a doubling of quiet sun power; enhancements of $100 - 300$\% are clearly seen, which is significantly larger than the observed halo values \citep{hindman1998, jain2002,schunker2010,rajaguru2012} of between 40-60\% and indeed the values achieved in 2D simulations \citep{khomenko2009,zharkov2013}. In our 3D simulations, where we have employed a simple gaussian wave source however, the magnitude is less of a concern than the qualitative behaviour of the enhancement.\\
We are primarily interested in whether mode conversion and the refraction of high frequency waves is the source of the halo enhancement. It is therefore necessary to ascertain whether the enhancements we see are true halos or simply an aberration caused by the magnetically modified atmosphere. In order to determine this we have run another simulation which we shall term the thermal case. The thermal simulation is performed with our $1.4$ kG sunspot atmosphere as normal, however the background field itself is removed everywhere when the simulation begins. In this instance the atmosphere is not really in magneto-hydrostatic equilibrium, however it tells us if the enhancement is caused by wave-field interactions (as it should be if mode conversion is the cause); If the field is the cause of the halo we expect to see no enhancement in the thermal case.\\
Figure \ref{sp2} shows the comparison between $P_{6.5}$ and $P_{\text{ther}-6.5}$, where where $P_{\text{ther}}$ is the analogue to $P$, denoting the time averaged power enhancement at every spatial point for the thermal case.\\
Clearly the thermal simulation yields no meaningful power enhancement, and so we conclude that the interaction between field and wave is key to the enhancement seen in this simulation. \\
To expand upon these simple findings we next show $P_{3.5}$, $P_{6.5}$ and $P_{\text{ther}-6.5}$ as functions of $x$ and $z$ in Figure \ref{pxz1}.
Figure \ref{pxz1} is simply the 2D version of Figures \ref{sp1} and \ref{sp2}, displayed for heights from $z=0$ to $z=1$ Mm above the surface, once again on the near side of the sunspot. \\
One can see from panel a) that the behaviour of $P$ is quite complex. Significantly, the halo is correlated with the $a=c$ layer, manifesting below it and spreading radially outwards with height, agreeing with observations \citep{rajaguru2012,schunker2011} and the prediction of \citet{khomenko2009}. There are two lobes of larger power separated by a region of weaker - but still significant - enhancement. There is of course no halo in either the 3.5 mHz or the thermal cases. \\
We next test the validity of our original assertion that the acoustic halo is generated by downwards turning fast waves. If this is the case, then if fast waves are prevented from returning from above the $a=c$ layer, the halo should disappear. We proceeded by performing 3 additional simulations to the ones already discussed. These simulations are identical to our primary 2-hour simulation (from which we have calculated $P$) in every way except that we enforce successively lower Alfv\'en speed limits of 50, 20 and 7 km/s on the atmosphere in the manner described in section 2. A maximum Alfv\'en speed of 7 km/s is just above the value at which $a=c$; therefore upcoming fast acoustic waves which undergo mode conversion to fast magentic waves are never able to achieve the condition for refraction, that $a=\omega/k_{h}$ (as $a$ is constant above this height), and so cannot return downwards and contribue to a halo. \\
Results of these simulations, showing the 6.5 mHz power enhancement, $P_{6.5}$, presented in the same manner as Figure \ref{pxz1} are displayed in Figure \ref{pxz2} below.

\begin{figure*}
\centering
    \includegraphics[width=0.99\textwidth,trim=6cm 0cm 6cm 1cm]{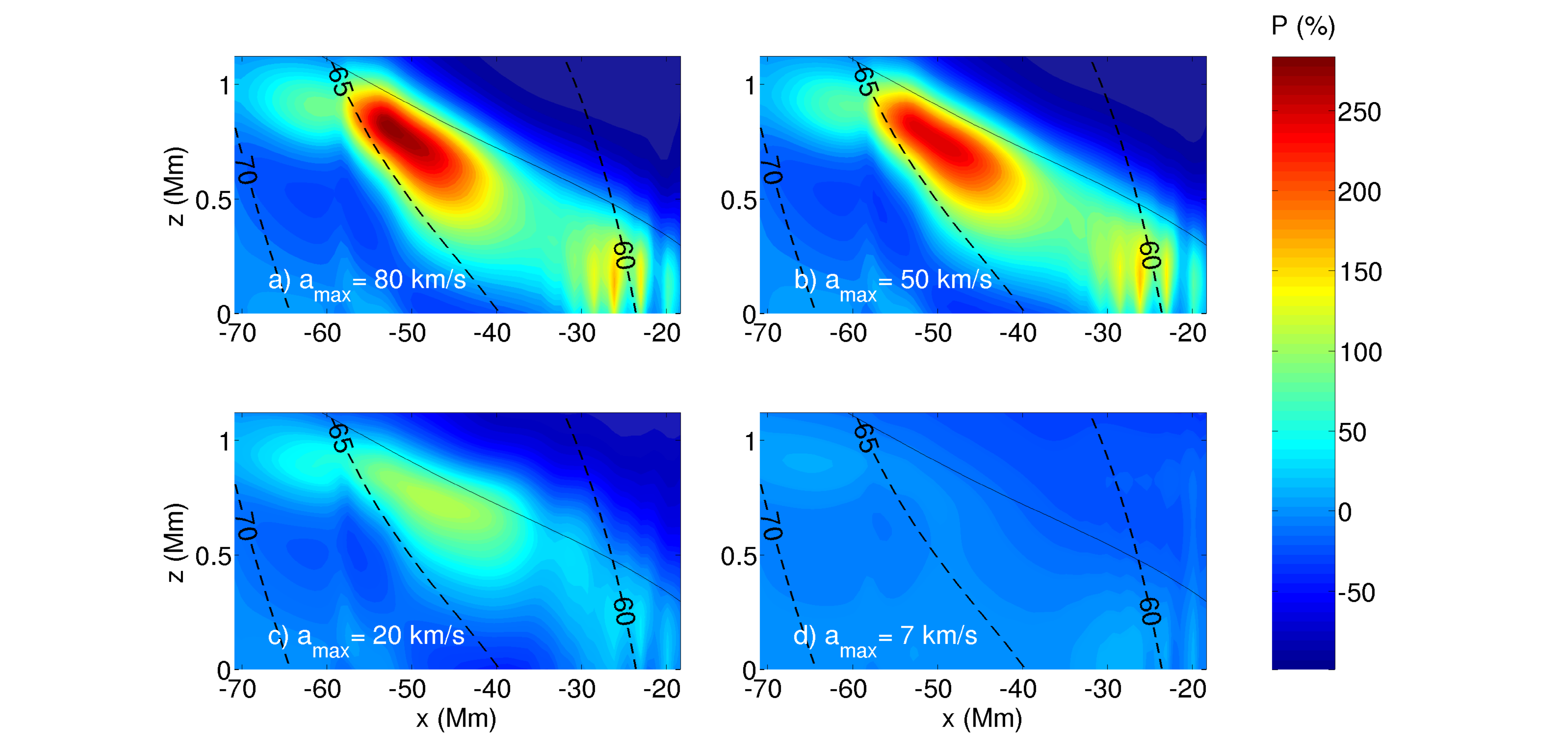}
    \caption{$P_{6.5}$ for the simulations with the background $a$ limited at a) 80, b) 50, c) 20 and d) 7 km/s.}\label{pxz2}
\end{figure*}

\begin{figure*}
 \centering
  \includegraphics[width=0.99\textwidth,trim=6cm 0cm 6cm 1cm]{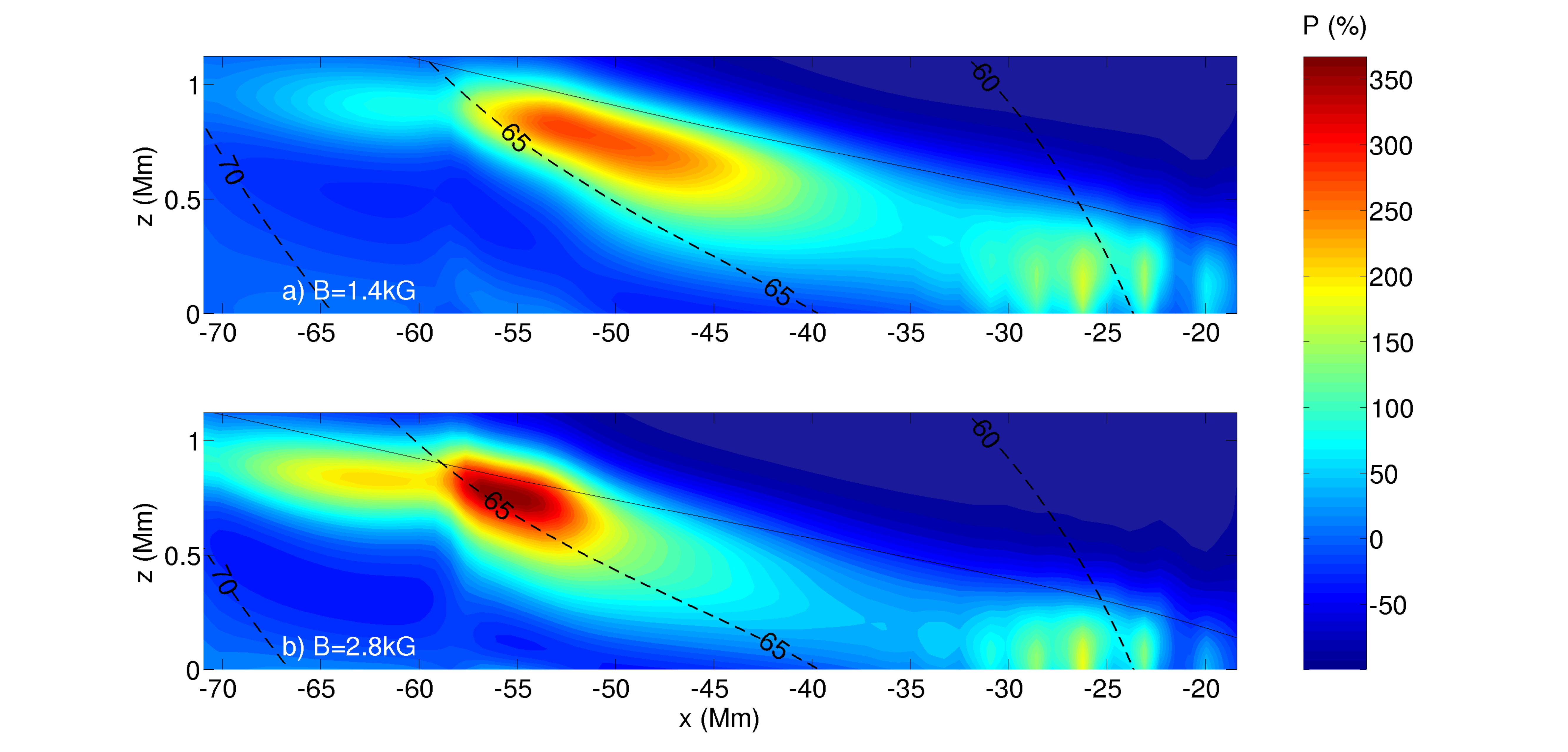}
    \caption{$P_{6.5}$ for the simulations with peak surface fields of a) 1.4 kG and b) 2.8 kG. The background $a$ is limited at 80 km/s in both cases.}\label{pxz3}
\end{figure*}

Panel d) clearly shows the complete reliance of the halo on the effects of the overlying magnetically dominated $a>c$ atmosphere.  To be clear, the atmospheres used for the 4 simulations are all identical below the $a=c$ layer. For d) the atmosphere is modified above this point such that waves cannot reflect and refract. The lack of any enhancement indicates that the halo is manifested purely as a result of waves which have refracted and reflected downwards through this overlying $a>c$ atmosphere.\\
Further evidence for the mode conversion hypothesis is presented in Figure \ref{pxz3}, which shows $P_{6.5}$ for the standard $B=1.4$ kG case (top panel) and for a simulation where everything was kept identical except the peak field strength of the sunspot was doubled to around $2.8$ kG (bottom panel).

\begin{figure*}
 \centering
  \includegraphics[width=0.99\textwidth,trim=6cm 0cm 6cm 1cm]{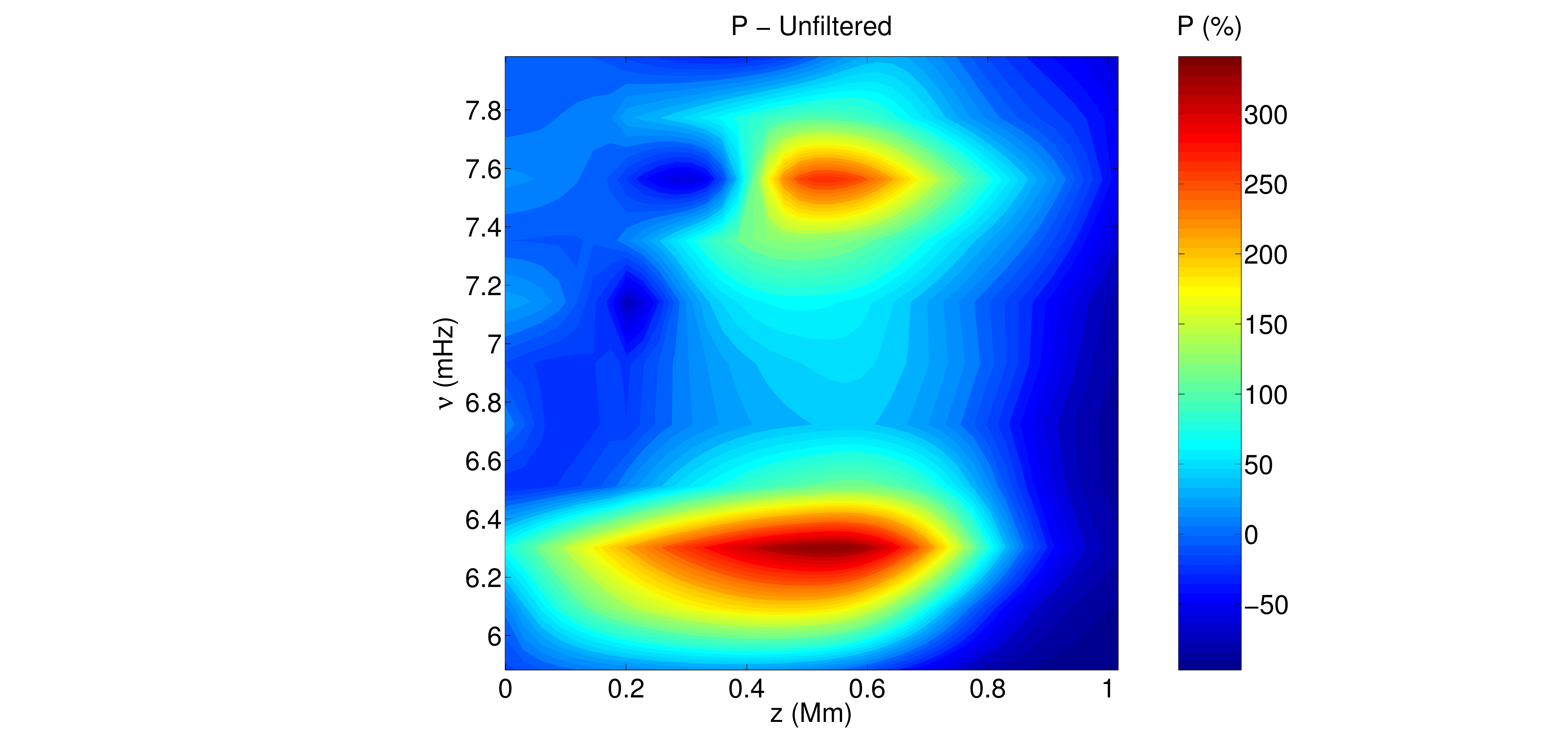}
   \caption{Unfiltered acoustic power $P$ as a function of height, averaged over the points from $x=-60$ Mm to $x=-20$ Mm in Figure \ref{pxz2}.}\label{pfz1}
\end{figure*}

In the case with the stronger 2.8 kG field, the halo is more spatially localised and the magnitude has increased by around 25\%. The greater field strength has focused the fast waves into a more confined region due to the lower fast wave reflection height. The halo itself has also dropped in height, corresponding to the lower $a=c$ layer present in the stronger sunspot atmosphere. \\
Observations of acoustic halo power also exhibit quite a clear spectral behaviour. The power appears to peak at higher frequency for greater heights of $z$ between 140 and 400 km and exhibits a dual lobe structure with peaks at around 6 and 8 mHz \citep{rajaguru2012}.
Shown in Figure \ref{pfz1} is the unfiltered acoustic power $P$ as a function of height (above $z=0$ km) and frequency.\\

The observed dual-lobe structure is evident in Figure \ref{pfz1}, peaking at 6.3 and 7.6 mHz for heights between 0 to 800 km above the photosphere. The second enhancement lobe at 7.6 mHz is more compact than the lower frequency lobe, only appearing at heights greater than 300 km. The fact that such a simple wave pulse simulation reproduces many of both the spatial and spectral properties of acoustic halos observed in the actual solar photosphere and chromosphere is certainly encouraging.

\section{DISCUSSION AND CONCLUSIONS}\label{sec:concl}
The results of our simple MHS sunspot wave propagation simulations show a marked power enhancement (by a factor between 1.5 and 4) with respect to quiet Sun values in the time averaged vertical component of velocity, $v_{z}$. The enhancement is present for relatively horizontal field (inclined 60 - 65 degrees from the vertical) which corresponds to weak field strengths of between 20-200 G. Spectrally, the enhancement exhibits twin peaks at approximately 6.3 mHz and 7.6 mHz, with the 7.6 mHz frequency peak manifesting slightly higher in the atmosphere than the lower frequency peak.\\
These characteristics (apart from the magnitude of the enhancement itself) match those determined observationally when acoustic power maps of halo regions have been analysed \citep{schunker2011,rajaguru2012}, indicating that we can, with reasonable certainty, refer to the enhancement as an acoustic halo. It is interesting that these features are brought out in such simple simulations with a wave source which differs so significantly from the bath of stochastically excited acoustic modes present in the real solar photosphere. \\
This fact suggests that the halo is a dynamic phenomenon brought about by the interaction of waves with the magnetic atmosphere, rather than any modification of the local acoustic oscillation frequency through granulation scale size shrinking \citep{jacoutot2008} or scattering effects \citep{hanasoge2009}.\\
The hypothesis with which we set out to investigate is that the acoustic halo is formed as extra energy is deposited into observable regions (in the photosphere and chromosphere) by downwards travelling fast waves which have refracted and reflected at the fast wave turning height. The idea was suggested and explored initially by \citet{khomenko2009}.\\
In the quiet Sun, an upwards travelling high frequency (non trapped) acoustic wave will continue to propagate upwards and out of the local area, taking its energy with it. However if there is a magnetic field present in the gravitationally stratified atmoshpere, wave energy will branch into the fast magnetic and the slow acoustic modes at around the $a=c$ equipartition region. The slow acoustic waves in this case are the modes which take energy away along the field lines. The fast magnetic waves will continue to travel upwards while refracting until the condition for reflection, that $\omega / k_{h} = a$ is met. It is these waves that would then be responsible for the excess energy.  
The fact that the power enhancement so closely correlates wih the $a=c$ layer (figures \ref{pxz1} - \ref{pxz3}) supports this hypothesis.\\
Furthermore, the halo magnitude scales with the value at which we cap the Alfv\'en speed, meaning that as we allow progressively more waves to return from the $a>c$ atmosphere, the halo becomes more apparent; When we do not permit upcoming waves to return downwards from above the $a=c$ layer, the halo disappears completely. This is the strongest piece of evidence in favour of the mode conversion mechanism. \\ 
The halo structure itself shows no evidence of any small-scale variations in magnitude, like the groupings of enhanced emission evident in egression power maps of active regions known as Glories \citep{braun1999,donea2000}. This is most likely due to the simplicity of the monolithic field structure and wave source which we have utilised. It is likely that the small bead-like glories require less idealised magnetic configurations with more fine structure than we have used here.\\ 
We have not undertaken any analysis of intensity halos in this study, owing to the simple nature of the wave pulse which we have used. The calculation of intensities requires a strong and continuous wave source, and so we intend to follow up this work by performing simulations with a more realistic stochastic and spatially homogeneous wave bath. In this manner we hope to produce the well observed intensity halo and also determine to what extent the velocity halo properties uncovered here are reproduced and/or extended.

\bibliographystyle{apj}        
\bibliography{database}
\end{document}